\documentclass[a4paper,11pt]{article}

\usepackage[utf8]{inputenc}
\usepackage[T1,T2A]{fontenc}
\usepackage{amsmath,amsfonts,amssymb}

 \usepackage{graphicx}
\usepackage{wrapfig}
\usepackage{subfig}
\usepackage{hyperref}
\usepackage[dvipsnames]{xcolor}

\begin{document}

\begin{center}
{\LARGE {\bf  Taming the dark photon production via a non-minimal coupling to gravity }}
\end{center}

\vspace{0.3cm}
\begin{center}
{\bf Oleg Lebedev$^{\; 1}$ and Jong-Hyun Yoon$^{\; 2 }$}
\\ \ \\
$^{1}$  \it{Department of Physics and Helsinki Institute of Physics,\\
  Gustaf H\"allstr\"omin katu 2a, FI-00014 Helsinki, Finland}\\
  $^{2}$  \it{Department of Physics and Institute of Quantum Systems (IQS), \\
  Chungnam National University, Daejeon 34134, Republic of Korea}\\
\end{center}

\vspace{0.7cm}
\noindent
{\bf Abstract.} Inflationary production of massive dark photons with  non-minimal couplings to gravity 
shows  surprising growth at large momenta. These  couplings appear in the effective low energy description
of a more fundamental theory. 
We find that the growth is absent in explicit gauge invariant UV-complete 
models. Such completions  are also free of ``ghost''  instabilities, which often appear in the effective models.
\\ \ \\

  The existence of a dark sector beyond the Standard Model  (SM) is motivated by various considerations including the 
  problem of dark matter and inflation.
In particular,   a massive vector (``Proca'')  field associated with the dark sector U(1) symmetry, dubbed a ``dark photon'', is an interesting 
dark matter candidate. It may be decoupled from the SM fields, but have non-minimal interactions with gravity \cite{Chernikov:1968zm}  which facilitate its
production in the Early Universe.  Including the lowest dimension terms, we obtain  the action  \cite{Novello:1979ik}
\begin{eqnarray}
S =  \int d^4x \sqrt{|g|} \left(-\frac{1}{4}F^{\mu\nu}F_{\mu\nu} + \frac{1}{2}m_A^2 g^{\mu\nu}A_\mu A_\nu 
  - \frac{1}{2}\xi_1 R g^{\mu\nu}A_\mu A_\nu - \frac{1}{2} \xi_2 R^{\mu\nu}A_\mu A_\nu\right)\;,
\label{eq:Sproca}
\end{eqnarray}
where $F_{\mu\nu} = \nabla_\mu A_\nu -  \nabla_\nu A_\mu$, $m_A$ is the dark photon mass, 
$\xi_i$ are dimensionless couplings, and   $R$ and $R_{\mu\nu}$ are the scalar curvature and the Ricci tensor, respectively.
In what follows, we restrict ourselves to the Friedmann space with the metric $g_{\mu\nu} = (1, -a^2,-a^2,-a^2)$, where $a=a(t)$
is a time-dependent scale factor.

In the Early Universe, the curvature terms are large which leads to efficient production of the vector quanta. The corresponding  study
has been performed in \cite{Capanelli:2024pzd}  with a surprising result that the production process
exhibits 
  a ``runaway''  feature. That is, for a certain range of $\xi_1$ and $\xi_2$, production of  high-momentum modes grows by many orders 
  of magnitude indicating instability of the system. The analysis of \cite{Capanelli:2024pzd}  has concluded  that there is no obvious solution   to the problem.
 Abnormalities in the behavior of a non-minimally coupled massive vector, including the ``ghost'' and tachyon features,  have also been observed in 
 \cite{Himmetoglu:2008zp}-\cite{Hell:2024xbv}. Issues with the model renormalizability have been discussed in \cite{Toms:2015fja,Buchbinder:2017zaa}.

In our work, we examine the problem from the viewpoint  of the ultra-violet (UV) completions of the non-minimal vector couplings. We find that 
the runaway behavior as well as other instabilities are absent in this case. Let us start by addressing the unitarity  problem of the model.

  \begin{figure}[h!]
    \centering
    \includegraphics[width=0.35\textwidth]{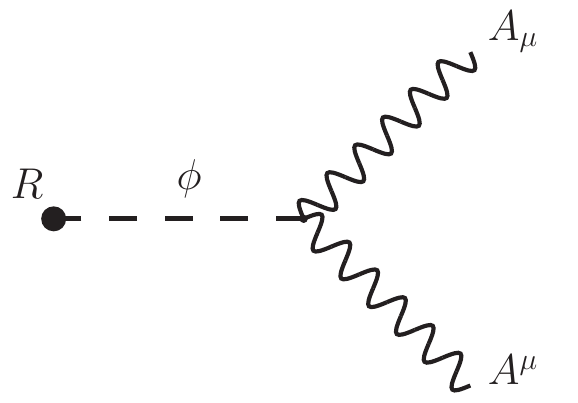}
     \includegraphics[width=0.35\textwidth]{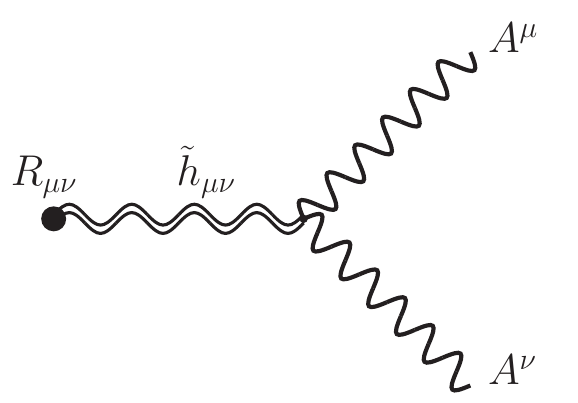}
    \caption{ Generation of the effective $\xi_i$ couplings in UV  complete theories. $\phi$ is a scalar and $\tilde h_{\mu\nu}$ is a Kaluza-Klein graviton.}
    \label{dia}
\end{figure}

{\bf \underline{Unitarity}.}
In the massless limit, 
the couplings $\xi_i$ violate gauge invariance. Hence, one expects $\xi_i \rightarrow 0$ as $m_A \rightarrow 0$.
This can also be seen from unitarity considerations. Indeed, the non-minimal couplings induce the vector scattering into gravitons $G$,
$$  AA \rightarrow GG  \;, $$
whose amplitude  ${\cal A}$  grows indefinitely with energy $E$. At high energies, the longitudinal vector components scale as $E/m_A$, which, together with the 
$E^2$--factor at the vertex, yields
\begin{equation}
|{\cal A}| \propto | \xi_{1,2} | \; {E^4 \over m_A^2 M_{\rm Pl}^2} \;,
\end{equation}
to lowest order in $|\xi_i| \ll 1$. 
This can lead to a scattering probability exceeding unity.
Therefore, perturbative unitarity breaks down at momenta of order
\begin{equation}
p_{\rm max } \lesssim {\sqrt{m_A M_{\rm Pl}}   \over   | \xi_{1,2} |^{1/4}  } \;,
\label{pmax}
\end{equation}
which represents the cutoff of the theory. The bound is independent of the specifics of the vector mass generation   and applies to both the Higgs and Stueckelberg mechanisms.
Unitarity considerations generally impose significant constraints on effective models with massive vectors (see e.g.\,\cite{Lebedev:2011iq}).

This result for the $\xi_1$ coupling can also be obtained in the Einstein frame, where the non-minimal coupling has been eliminated by the 
metric rescaling $g_{\mu\nu} = \left( 1- {\xi_1 \over M_{\rm Pl}^2}\, A_\rho A^\rho    \right) \tilde g_{\mu\nu}$ \cite{Salopek:1988qh} in favor of higher dimensional operators
containing $A_\mu$. To lowest order in $\xi_1$, this generates ${\xi_1 \over M_{\rm Pl}^2 } \; m_A^2\, (A_\mu A^\mu)^2$ and induces $AA \rightarrow AA$
scattering with the unitarity cutoff (\ref{pmax}).

The range of validity of the theory shrinks to zero as $m_A \rightarrow 0$. However, if $|\xi_i| \propto m_A^2$, it remains meaningful in the massless limit,
which is what we find in explicit UV completions of the model. The result also applies to curved spaces as long as the relevant momenta are above the inverse curvature radius.

 Clearly, the theory with the non-minimal couplings is effective and must be UV-completed, whether the vector mass is Higgs-- or Stueckelberg--generated. 
 The couplings correspond to effective ``form-factors'' obtained by integrating out heavy states. 
 Such form-factors must be constant within the energy range of the problem, 
which brings in further constraints. In particular, the particle production calculations assume that $m_A$ and $\xi_i$ 
remain constant for the characteristic momenta between zero and the Hubble scale, or even above the Hubble scale.
We find that this imposes a $stronger$  constraint on the size of the effective couplings than the unitarity considerations do.

In what follows, we  consider UV completions of the non-minimal vector couplings to gravity. We focus on the vector mass generation due to the Higgs mechanism, in which case healthy
 models valid up to the Planck scale can be constructed.

{\bf \underline{Scalar coupling}.} Consider a complex scalar $\Phi$ charged under a gauged U(1) and possessing a non-minimal coupling to gravity,
 \begin{equation}
 {\cal L}_{\rm sc} 
 = \overline{D_\mu  \Phi} \, D^\mu \Phi - {1\over 2} \xi \,R\, |\Phi|^2 - V(\Phi)\;.
 \end{equation}
Here $D_\mu = \partial_\mu -igA_\mu $ with $g$ being the gauge coupling. 
 This theory is  well-behaved and   only limited by the gravitational cutoff around the Planck scale  unless 
 $\xi$ is very large \cite{Barbon:2009ya}.

 When $\Phi$ develops a vacuum expectation value  (VEV) $v/\sqrt{2}$, the gauge field attains mass
 $m_A = g v$. Apart from the massive vector, the particle spectrum  contains a real scalar  
 $\phi   = \sqrt{2} \, |\Phi| -v $    with bare mass $m_s$ determined by the potential $V(\Phi)$, which also receives 
 a cosmologically induced mass-squared contribution $\xi R/2$.
 If $m_s$ is the heaviest scale in the problem, the scalar can be integrated out,  which produces an effective $\xi_1$ coupling (Fig.\,\ref{dia}, left). In the Friedmann Universe, this corresponds to 
shrinking the scalar propagator $\left(  \partial_\mu \partial^\mu +3H\partial_0  +m_s^2 + \xi R/2 \right)^{-1}$ to a point.
We then find
 \begin{equation}
 {\cal L}_{\xi_1} 
 =  -{1\over 2} \xi \, {m_A^2 \over m_s^2}\, R\, A_\mu A^\mu ~~,~~ \xi_1 = \xi  \, {m_A^2 \over m_s^2}\;.
 \label{eff-xi}
 \end{equation}
  The effective field theory description requires that  the momenta be below $m_s$ and that $\xi R/2 $ can be neglected,
    \begin{equation}
 p \ll  p_{\rm max} \sim \sqrt{\xi\over \xi_1} m_A~~,~~  |\xi  R| /2 \ll m_s^2 \;.
 \label{eff-xi}
  \end{equation}
  At larger
  momenta,  $\xi_1$ is no longer constant and the corresponding amplitude scales as $1/p^2$ at high energies.
  The requirement $ |\xi  R| /2 \ll m_s^2$ is equally important since it ensures that
  the scalar VEV and $m_A$ remain constant as the Hubble rate changes, in particular, over  the vector production period.
    During inflation with Hubble rate $H$, $R=-12H^2$, hence 
it  sets the constraint
   \begin{equation}
 |\xi_1| \ll {1\over 6} {m_A^2\over H^2}\;.
 \label{xi1bound}
  \end{equation}
  If this bound is violated, $m_A$ and $\xi_1$ become time-dependent. Therefore, only a small non-minimal coupling is allowed. A similar conclusion has been reached in Ref.\,\cite{Cyncynates:2024yxm} for the  flat space case using  dimensional analysis. This result can  trivially  be generalized to 
  models with multiple scalars.

We note that Ref.\,\cite{Capanelli:2024pzd} has considered a Higgs Abelian extension of the vector non-minimal coupling with $\xi=0$ at tree level. Since gravity preserves 
gauge symmetry, the loop effects amount to generating gauge invariant corrections to the tree level Lagrangian and  
a small non-zero $\xi$ is thus induced. Then, our procedure of integrating out a heavy scalar applies and  the same conclusion is reached. The difference from the treatment  of Ref.\,\cite{Capanelli:2024pzd}  is that we keep the external momentum (and the curvature) dependence in the propagators, which is required by the unitarity and EFT considerations. In particular, the non-minimal vector coupling must vanish at large momentum transfer. Our Eq.\,\ref{xi1bound} applies to the special case  $\xi\vert_{\rm tree}=0$ as well.

{\bf \underline{Tensor  coupling}.}  $\xi_2$ is specific to vector fields and has no scalar analog, which makes it less straightforward to generate.  
The coupling structure suggests that it can be obtained by integrating out a symmetric tensor field. The prime, well-behaved candidate for this role is the Kaluza-Klein (KK) graviton,
which appears in models with extra dimensions.
Consider the minimal case of a 5-d space with the 5th dimension being compactified on a circle of radius $r/(2\pi)$. Suppose that gravity propagates in the bulk, while the matter fields
are confined to a 3-d brane \cite{Arkani-Hamed:1998sfv}. The corresponding action is
\begin{equation}
S= \int d^5 x \, {1\over \hat \kappa^2} \sqrt{|\hat g|} \hat R  + \int d^5 x  \sqrt{|\hat g|}\, {\cal L}_{\rm mat}\, \delta (x_5) \;,
\end{equation} 
where the hatted quantities are  5-dimensional, $\hat g$ is the metric determinant,  ${\cal L}_{\rm mat}$ is the matter Lagrangian and $\hat \kappa^2 $ is the 5-d Newton constant multiplied by $16\pi$.
Since the translational invariance along the 5th coordinate is broken explicitly, one may also add a localized gravity term, which can be  induced by quantum corrections:
 \begin{equation}
S_{\rm loc}=\epsilon  \, \int d^5 x \, {1\over  \kappa^2} \sqrt{| \hat g|}\,   R  (\hat g_{\mu\nu})  \, \delta (x_5)\;,
\label{Palatini}
\end{equation} 
where $|\epsilon| \ll 1$ and $1/\kappa^2 $ is the 4-d Newton constant multiplied by $16\pi$.  
Below the scale of the KK modes, we recover the Einstein gravity with 
$M_{\rm Pl}^2/2 = r/{\hat \kappa^2} +\epsilon /\kappa^2$ and a set of higher dimensional operators. By construction, the 4d Planck mass is dominated by the 5d-induced term $r/{\hat \kappa^2}$.

The Kaluza-Klein gravitons are obtained by expanding the metric $\hat g_{\mu\nu} = \eta_{\mu\nu} +  \kappa \, h_{\mu\nu}+...$, where all the indices are 4-dimensional,  $\kappa \simeq  \hat \kappa/\sqrt{r}$ and we omit the dilaton mode irrelevant to our discussion.
Following the analysis and conventions of \cite{Han:1998sg}, we expand $h_{\mu\nu}$
  in the Fourier modes,   
\begin{equation}
h_{\mu\nu}(x, x_5)= \sum_n  h_{\mu\nu}^n(x)\, \exp\left( i{2\pi n x_5 \over r}    \right)\;,
\end{equation}
 where $n$ enumerates the Kaluza-Klein modes with mass $m_n = 2\pi n/r$. The canonical normalization of the graviton modes requires a shift $h_{\mu\nu} \rightarrow \tilde h_{\mu\nu}  + \alpha_{\mu\nu}$, which involves the extra-dimensional components 
 of the graviton.  
 The canonically normalized KK graviton couples to the energy momentum tensor of matter fields the same way the 4d graviton does, $\kappa \tilde h^{\mu\nu,n}T_{\mu\nu}$, up to the $\epsilon$-correction to the Planck mass.  In particular, the tensor coupling to the complex scalar charged under U(1)  is given by 
   $ \Delta {\cal L}_{\rm mat}=- \kappa \, \tilde h^{\mu\nu, n} \, \overline{D_\mu \Phi}\, D_\nu \Phi $  \cite{Han:1998sg}.
 Since $\delta R / \delta g_{\mu\nu} = R^{\mu\nu}$ up to the boundary terms, 
 the localized curvature term (\ref{Palatini}) leads to the tensor coupling ${\epsilon \over \kappa} \sqrt{| g|}\,   R_{\mu\nu}   \; \tilde h^{\mu\nu, n}  $. Using the KK graviton propagator 
 $i\Delta_{\mu\nu,\rho\sigma}= {i\over {p^2-m_n^2}}\,(\eta_{\mu\rho} \eta_{\nu\sigma} + \eta_{\mu\sigma} \eta_{\nu\rho} - {2\over 3} \eta_{\mu\nu} \eta_{\rho\sigma}) + {\cal O} (p^2/m_n^2)$, we 
 can integrate out the massive graviton, obtaining the effective 4d term
 \begin{equation}
 \Delta {\cal L} = - {\epsilon \over m_n^2} \,   R_{\mu\nu}  \,\left(  \overline{D^\mu \Phi}\, D^\nu \Phi  + \overline{D^\nu \Phi}\, D^\mu \Phi\    \right)\;,
 \end{equation}
 as well as higher derivative interactions with different tensor  structures and the scalar curvature term.

 When the scalar develops a VEV $v$,  the vector field attains mass  and we
 obtain the  effective  non-minimal coupling of the vector to gravity (Fig.\,\ref{dia}, right). The sum over the KK modes converges and the result is dominated by the lightest  mode,
    \begin{equation}
   {\cal L}_{\xi_2} \simeq   -   \epsilon \, {m_A^2 \over m_1^2}    \, R_{\mu\nu}A^\mu A^\nu ~~,~~         \xi_2 =  2\epsilon \, {m_A^2 \over m_1^2}\;.
 \end{equation}
   The effective theory breaks down at $p \sim m_1$, which implies 
 \begin{equation}
 p_{\rm max} \lesssim {m_A \over \sqrt{|\xi_2|}} \;.
 \end{equation}
 At higher energies, the non-minimal coupling is no longer constant and higher derivative couplings become  important. The KK graviton exchange  at energies above the cutoff 
 leads to the formfactor 
   behavior $\xi_2  \propto 1/p^2$.    
   We note that the $\xi_1$ coupling of  similar size   is also induced by the KK graviton exchange,   $\xi_1 \sim {\cal O}(\xi_2)$. 
   The above  conclusion    equally applies to the situation when 4d gravity is dominated by the localized term. Indeed, in this case $\epsilon \simeq 1$ and $\epsilon /\kappa^2 \gg r/{\hat \kappa^2}$, such that the KK modes couple weaker than the zero modes do.
This suppression factor plays the role of $\epsilon$ above, leading to a similar result for $\xi_i$.

This analysis is done in the linearized gravity approximation, yet the results are  general. Indeed, general covariance fixes the structure of the effective coupling, while integrating out the tensor field is justified when it is heavier than
the other relevant scales of the problem, e.g. the Hubble scale, $m_1\gg H$.  At  very large momenta $p \gg H$, the flat space results apply and $\xi_2  \propto 1/p^2$.
 As in the scalar coupling case, we find
  \begin{equation}
|\xi_{1,2}| \ll  {m_A^2  \over H^2} \;.
 \end{equation}

   The above  theory is only limited by the 
  gravitational cut-off given by the 5-d analog of the Planck mass, $(1/\hat \kappa^2)^{1/3}$ \cite{Arkani-Hamed:1998sfv}, and hence represents a legitimate UV-completion for the effective theory 
  with a  non-minimal gravity coupling.

 {\bf \underline{Generalization}.} 
 These examples show a simple  pattern. As is clear from general considerations, the non-minimal vector coupling to gravity cannot be  a fundamental quantity and is obtained in the low energy limit by integrating out heavy states. 
  In the UV complete theory, it  tends to zero at high energies as required by unitarity and vanishes in the massless limit by gauge invariance. The simplest Ansatz  for the coupling  satisfying these requirements has the form of a heavy particle propagator,
  up to an order one coefficient,
  \begin{equation}
  \xi_{i} \propto {m_A^2  \over p^2 -M^2 + {\cal O}(H^2)} \;,
  \label{ansatz}
  \end{equation}
  where $p$ is the relevant momentum scale, $M$ is a large mass scale and 
  the Hubble scale corrections appear in the Friedmann Universe. In this case, the unitarity cutoff is the Planck scale.
  
  This Ansatz has an important consequence. If one requires $\xi_i$ to be constant in a wide energy range up to  the Hubble scale and above, the mass scale $M$
  must be far greater than  $H$. As a result, the low energy value of $\xi_i$ is bounded: 
  \begin{equation}
  |\xi_i | \ll m_A^2/H^2 \;. 
  \label{gen-bound}
 \end{equation}
 The momentum cutoff of the effective theory is of order $m_A/ \sqrt{|\xi_i|}$. Analogous results apply to the case of multiple heavy particle propagators in (\ref{ansatz}).
 
 Eq.\,\ref{ansatz} can also be used in (nearly) flat space to study interactions of light dark photons. When $H \sim 0$, the EFT description is valid up to the momenta close to  the heavy particle threshold $M$. For example, MeV scale dark photons with a substantial  non-minimal coupling to gravity represent a viable option at low energies. However,  this EFT breaks down in the Early Universe, when the particle momenta become large, and particle production cannot be studied in this approximation.
 
 {\bf \underline{Particle production}.}  The Proca field is produced efficiently by inflation \cite{Graham:2015rva}-\cite{Grzadkowski:2024oaf}.  Its abundance can be computed using the standard particle production  techniques. 
 The field is decomposed in the spacial Fourier modes as $A^\mu =\int {d^3 \boldsymbol{k} \over (2\pi)^3} A^\mu_{\boldsymbol{k}} (t) e^{i \boldsymbol{k}\cdot \boldsymbol{x}}$.
 The 
 temporal component is non-dynamical and can be integrated out. The remaining vector components are split into the transverse part, $\boldsymbol{k} \cdot {\vec{A^T_{\boldsymbol{k}}}}=0$,
and the longitudinal part,   $\boldsymbol{k} \cdot {\vec{A^L_{\boldsymbol{k}}}}= k\, A^L_{\boldsymbol{k}} $, where $k\equiv |\boldsymbol{k}|$.
 The transverse components ${\vec{A^T_{\boldsymbol{k}}}}$  behave as standard massive scalars, while the action for the longitudinal part ${A^L_{\boldsymbol{k}}}$ 
 in terms of the conformal time $d\eta= dt /a$
 reads  \cite{Kolb:2020fwh}
  \begin{eqnarray}
S^L & = &\displaystyle\int d\eta \int \dfrac{d^3 \boldsymbol{k}}{(2 \pi)^3} \bigg[\frac{1}{2} \dfrac{a^2 m_t^2}{k^2+a^2 m_t^2} \left|\partial_0 A_{\boldsymbol{k}}^L\right|^2  
  -\frac{1}{2} a^2 m_x^2\left|A_{\boldsymbol{k}}^L\right|^2\bigg] \;,
\end{eqnarray}
 where
  \begin{eqnarray}
  &&  m^2_t = m_A^2 - \xi_1R - \tfrac{1}{2}\xi_2 R - 3 \xi_2 H^2\, , \label{eq:mt}\\
   && m^2_x = m_A^2 - \xi_1R - \tfrac{1}{6}\xi_2R + \xi_2H^2\, .\label{eq:mx}
\end{eqnarray}
 The kinetic term exhibits  a surprising abnormality: it can turn negative for light enough vectors, leading to a ``ghost''-type  instability. This is indicative of the effective field theory description being problematic
 at high energies, i.e. when $R$ and/or $H$ are  large relative to the particle mass. Furthermore, even if one judiciously chooses the parameters as to avoid the ghost feature, the mass term  $m_x^2$ 
 can still be negative for cosmologically long time-scales, indicating a tachyonic instability. 
 
 These problems are absent in the UV-completions. Indeed, requiring $\xi_i$ to be constant in a wide energy range up to the Hubble scale, Eq.\,\ref{gen-bound} implies that the couplings are small  and 
 $m_A^2 \gg |\xi_i R|, |\xi_i| H^2$. Thus, both $m^2_t$ and $m^2_x$ are positive.
 
To compute particle production for given $\xi_i$, it is convenient to canonically normalize  $A_{\boldsymbol{k}}^L$ assuming that $m^2_t >0$. Introducing $\chi_{\boldsymbol{k}}(\eta) $ according to
$A_{\boldsymbol{k}}^L(\eta)=\kappa_k(\eta) \chi_{\boldsymbol{k}}(\eta)  $ with $ \kappa^2_k(\eta)=\frac{k^2+a^2 m_t^2}{a^2 m_t^2} \, ,$ one finds the Lagrangian 
$ {\cal L}_\chi= \tfrac{1}{2}\left|\partial_\eta \chi_k\right|^2-\tfrac{1}{2} \omega_k^2|\chi_k|^2 $, where $ \omega_k^2$ can be negative depending on the sign of $m^2_x$.
 The full expression for $ \omega_k^2$ can be found in \cite{Capanelli:2024rlk}. Its important feature is that it contains a  term that grows with the momentum as $k^2 m_x^2/m_t^2$ and can lead to strong tachyonic
 particle production for $m^2_x<0$.
 Solving the equation of motion for $\chi_k (\eta)$, one computes the 
    comoving number density  according to  $n_k=k^3|\beta_k|^2/2\pi^2 $ with $  |\beta_k|^2 = \omega_k|\chi_k|^2/2 + |\partial_\eta \chi_k|^2/2\omega_k-1/2$  \cite{Capanelli:2024rlk}.

    \begin{figure}[h!]
    \centering
    \includegraphics[width=0.55\textwidth]{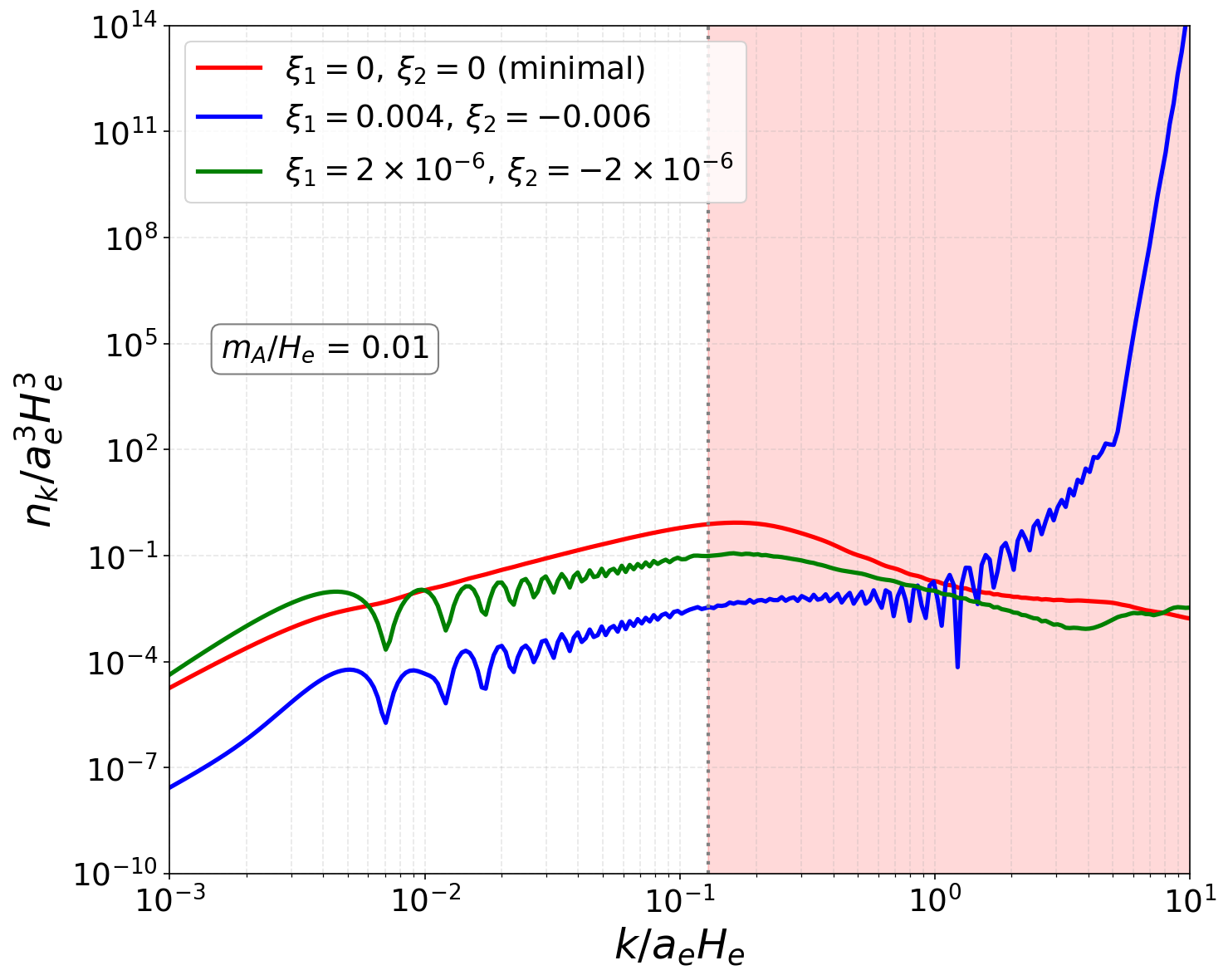}
    \caption{ The spectral density of the produced particles as a function of the momentum.  In the shaded area, the effective theory breaks down for the parameters of the blue curve ($\xi_1=0.004, \xi_2 =-0.006$),
   which requires the momentum cutoff $m_A/\sqrt{|\xi_2|}  \simeq 0.1 H_e$.}
    \label{fig}
\end{figure}

    Fig.2 shows the results of our numerical analysis.  The particle density is normalized as  $n_k /(a_e^3 H_e^3)$, where $a_e$ and $H_e$ are the scale factor and the Hubble rate at the end of inflation, respectively.
    The comoving momentum $k$ is related to the physical momentum by $k/a = p$.
    Following Ref.\,\cite{Capanelli:2024pzd}, we assume that inflation is succeeded  by a matter dominated epoch with the average equation of state $w=0$ such that 
  $R=-3H^2$. 
    We find reasonably good consistency with the results  of Ref.\,\cite{Capanelli:2024pzd}, within the uncertainties of the assumed initial conditions and numerical implementation.  
    In particular, we observe the surge in particle production at $k/(a_e H_e) \gtrsim 1$ for $\xi_1=0.004, \xi_2 =-0.006$. This feature is absent for smaller couplings.

In the UV completions, the sharp increase in particle production appears beyond the cutoff of the effective theory. Indeed, the momentum cutoff is $m_A/\sqrt{|\xi_i|}$, while for larger physical momenta
$\xi_i$ can no longer be treated as constant (shaded area in Fig.2).
If one requires the effective couplings to be constant up to $k/a =10 H$, the magnitude of the couplings is bounded by $ m_A^2/(10H)^2 \sim 10^{-6}$ for the parameter choice in the figure. 
The green curve shows that particle production remains rather mild in this case and close to that for the minimally coupled vector.
Hence, no ``runaway'' particle production occurs in the UV-complete models.  

To summarize, the non-minimal couplings of the Proca field to gravity are meaningful within effective field theory. This theory has a unitarity cutoff that approaches zero in the massless limit,
which necessitates  model extension at high energies.
We have constructed UV completions of the effective theory that are  limited by the Planckian physics only.  In these completions, the non-minimal couplings are obtained by integrating out heavy spin-0 and spin-2 fields
such that $\xi_i$ behave as formfactors. The resulting effective theory is limited by the mass scale of these fields, beyond which the formfactors are no longer constant. If one requires the non-minimal couplings
to stay constant up to the Hubble scale, their size is constrained to be small. As a result, the theory is ghost- and tachyon-free, and no runaway vector production is allowed.

General unitarity and gauge invariance arguments show that the above features are generic  in  UV completions and the non-minimal couplings are given by the Ansatz of Eq.\,\ref{ansatz}. 

\noindent
{\bf Acknowledgements.} OL is grateful to Adam Falkowski for useful communications. 
This work was supported by the National Research Foundation of Korea (NRF) grant funded by the Korea government (MSIT) (RS-2024-00356960, RS-2025-00559197) and by the National Supercomputing Center with supercomputing resources including technical support (KSC-2025-CRE-0282).

\end{document}